\documentclass[aps,prb,twocolumn]{revtex4}

\usepackage{bm}
\usepackage{dsfont}
\usepackage{graphicx}
\usepackage{amsmath}
\usepackage{amssymb}

\renewcommand{\H}{{\cal{H}}}
\newcommand{\bp}{\bm{p}}
\newcommand{\bq}{\bm{q}}

\newcommand{\bA}{\bm{A}}
\newcommand{\bB}{\bm{B}}
\newcommand{\bO}{\bm{O}}
\newcommand{\bV}{\bm{V}}
\newcommand{\bE}{\bm{E}}
\newcommand{\bF}{\bm{F}}
\newcommand{\bG}{\bm{G}}
\newcommand{\bH}{\bm{H}}
\newcommand{\bK}{\bm{K}}
\newcommand{\bM}{\bm{M}}
\newcommand{\bX}{\bm{X}}
\newcommand{\bR}{\bm{R}}
\newcommand{\bLm}{\bm{\Lambda}}
\newcommand{\bOm}{\bm{\Omega}}
\newcommand{\tS}{\widetilde{S}}
\renewcommand{\aa}{\hat{a}}
\newcommand{\ad}{\hat{a}^\dagger}
\newcommand{\bb}{\hat{b}}
\newcommand{\bda}{\hat{b}^\dagger}
\newcommand{\hp}{\hat{p}}
\newcommand{\hq}{\hat{q}}
\newcommand{\bi}{{\bm{i}}}
\newcommand{\bd}{{\bm{d}}}
\newcommand{\bk}{{\bm{k}}}
\newcommand{\bpi}{{\bm{\pi}}}
\newcommand{\ave}[1]{\langle#1\rangle}
\newcommand{\bigave}[1]{\big\langle#1\big\rangle}

\newcommand{\dave}[1]{\langle\hspace{-0.9mm}\langle#1
                      \rangle\hspace{-0.9mm}\rangle}
\newcommand{\bigdave}[1]{\big\langle\hspace{-1.1mm}\big\langle#1
                         \big\rangle\hspace{-1.1mm}\big\rangle}

\begin{document}

\title{Quantum effects in nanosystems:\\
	   good reasons to use phase-space Weyl symbols}

\author{Ruggero Vaia}
\email{ruggero.vaia@isc.cnr.it}
\affiliation{Istituto dei Sistemi Complessi,
             Consiglio Nazionale delle Ricerche,
             via Madonna del Piano 10,
             I-50019 Sesto Fiorentino (FI), Italy}
\affiliation{Istituto Nazionale di Fisica Nucleare,
	         Sezione di Firenze,
             via G.~Sansone 1,
             I-50019 Sesto Fiorentino (FI), Italy}
\date{\today}

\begin{abstract}
Bogoliubov transformations have been successfully applied in several Condensed Matter contexts, e.g., in the theory of superconductors, superfluids, and antiferromagnets. These applications are based on bulk models where translation symmetry can be assumed, so that few degrees of freedom in Fourier space can be `diagonalized' separately, and in this way it is easy to find the approximate ground state and its excitations. As translation symmetry cannot be invoked when it comes about nanoscopic systems, the corresponding multidimensional Bogoliubov transformations are more complicated. For bosonic systems it is much simpler to proceed using phase-space variables, i.e., coordinates and momenta. Interactions can be accounted for by the self-consistent harmonic approximation, which is naturally developed using phase-space Weyl symbols. The spin-flop transition in a short antiferromagnetic chain is illustrated as an example. This approach, rarely used in the past, is expected to be generally useful to estimate quantum effects, e.g., on phase diagrams of ordered vs disordered phases.
\end{abstract}

\pacs{03.65.Sq,05.30.Jp,05.30.Rt,75.10.Pq}


\maketitle

\section{Introduction}

The purpose of this paper is to propose a simple approach for studying the ground state or the thermal state of a quantum system. It will be argued that it is often convenient to deal with the {\em phase-space} formalism in terms of generalized coordinates and momenta, rather than with the {\em bosonic} formalism using creation/annihilation operators. The self-consistent harmonic approximation~\cite{Hooton1955,Koehler1966,KleinH1972} (SCHA), which is based on the frequency renormalization arising from nonlinear interactions and aims at the best harmonic approximation of the density matrix of the nonlinear system, can indeed be well formulated by means of phase-space variables.

The treatment of interacting systems is indeed simpler by working with (real) $c$-numbers in the place of (Hermitian) operators, namely using their `Weyl symbols'. The main reason why Weyl ordering of quantum operators and the corresponding symbols~\cite{Weyl1927,Berezin1980} have been employed is in that they naturally relate quantum systems to their classical counterparts, which makes them very useful on the `semiclassical' side. A well-known example is the Wigner distribution function~\cite{Wigner1932,ImreORZ1967,HilleryCSW1984}, which is the Weyl symbol for the thermal-equilibrium density operator: it has the captivating property that it can be used in familiar classical formulas, e.g., the averages of observables (of course, with proper caveats). However, Weyl symbols can be helpful also on the `quantum' side, namely for problems that are usually faced in terms of bosonic operators. Since most often the starting point is a normal-ordered Hamiltonian, it is useful to recall~\cite{AgarwalW1970a,AgarwalW1970b,Lee1995} and derive (new) relations  connecting normal symbols with Weyl symbols.

Furthermore, the choice of working in the phase-space formalism is motivated by the observation that Bogoliubov transformations, which are ubiquitous when dealing with harmonic approximations, arise more naturally than in the boson-operator formalism. This is clearly evident even in the simplest implementation for a single degree of freedom, $\aa=(\hq{+}i\hp)/\sqrt2$, where the transformation
\begin{equation}
 \bb=\cosh\!\chi\,\aa+\sinh\!\chi\,\ad
\label{e.Bogol}
\end{equation}
(canonical, as $[\bb,\bda]=1$), puts into normal form Hamiltonians such as
\begin{eqnarray}
 \hat\H &=& \frac{\omega_0}2(\ad\aa{+}\aa\ad) +\frac{\gamma}2(\ad\ad+\aa\aa)
\nonumber\\
  &=& -E_0+\omega\,\bda\bb ~;
\end{eqnarray}
the parameter $\chi$ and the proper frequency $\omega$ are found by replacing~\eqref{e.Bogol} in the last line and finding two equations, $\omega\cosh2\chi=\omega_0$ and $\omega\sinh2\chi=\gamma$, whose solution  eventually yields $\tanh2\chi=\gamma/\omega_0$ and $\omega=\sqrt{\omega_0^2{-}\gamma^2}$. 
However, no algebraic calculations are needed in terms of phase-space operators: the proper frequency arises immediately, since
\begin{equation}
 \hat\H=\frac{\omega_0{-}\gamma}2\,\hp^2+\frac{\omega_0{+}\gamma}2\,\hq^2
\end{equation}
is just a harmonic oscillator, which makes also evident why it must be $|\gamma|<\omega_0$. When many degrees of freedom are involved, the phase-space formalism allows one to use ordinary linear canonical transformations in the place of multidimensional Bogoliubov transformations: the latter involve `para-unitary' matrices~\cite{Colpa1978, Tsallis1978} and are definitely more complicated, although equivalent~\cite{Broadbridge1979}. Therefore, the phase-space approach is markedly convenient for nanosystems, since translation symmetry cannot be invoked in order to reduce the required transformations within Fourier channels with few degrees of freedom, as it is commonly made for the corresponding bulk counterparts. 

Of course, if in the Hamiltonian momenta and coordinates are separated in a kinetic and a potential part ({\em standard} Hamiltonian), working in phase space seems to be obvious, since linear canonical transformations reduce to orthogonal transformations; nevertheless, for such systems many textbooks and papers use the bosonic approach and reformulate the Hamiltonian in terms of creation/annihilation operators. As remarked above, this makes little difference when translation symmetry holds, as, e.g., in the study of crystal-lattice vibrations. However, the category also includes finite and/or disordered assemblies of atoms or molecules, and for them the phase-space SCHA can be more conveniently used; a typical example where it constitutes a viable alternative to heavier quantum numerical approaches~\cite{LeeB2001,PerdomoDDRA2012} is the task of finding the ground-state of a molecular cluster (such as a protein model): the SCHA renormalization of classical relative-minimum configurations permits to recognize the quantum energy minimum, namely the true ground state.
On the other hand, several Condensed Matter Hamiltonians are formulated from the very beginning in terms of creation/annihilation operators. Among them, those  describing cold atoms in optical lattices, which have become a training ground where the interactions can be engineered to simulate old and new quantum mechanical systems~\cite{BlochDZ2008,BlochDN2012}; for bosonic degrees of freedom, many variants of the Bose-Hubbard model can be realized, whose study can be performed by means of phase-space Weyl symbols and SCHA.
The same approach is particularly useful for magnetic systems, i.e., assemblies of spins interacting in whatever way~\cite{Mattis1985,Yoshida1996}, from short chains of spins in ring-shaped molecules, to short open magnetic chains that can result from impurities in quasi-one-dimensional magnetic materials. An antiferromagnetic model is chosen in this paper in order to exemplify the application of the method.
Extended quantum system often show {\em quantum} phase transitions, corresponding to qualitative ground-state changes that occur while varying a parameter of the Hamiltonian. It is to be remarked that in several cases, as in the paradigmatic transverse-field Ising model~\cite{Sachdev1999}, these changes are related to a transition between different minima of the corresponding classical system, so the quantum phase diagram can be obtained in a semiclassical way~\cite{CTVV2007prb}.
Also frustrated systems can be faced; e.g, in Ref.~\onlinecite{ColettaZM2013} one finds an example of `quantum stabilization' of a classically unstable phase, namely the magnetization plateau in the phase diagram of the 2D $J_1$-$J_2$ frustrated antiferromagnet: the theory follows a creation/annihilation-operator approach on few Fourier channels for a bulk system: in the case of a finite sample, the approach proposed here would be equivalent, but more convenient.

In Section~\ref{s.normal-weyl} the definition of normal and Weyl symbols is recalled, along with some properties and useful formulas. Then, in Section~\ref{s.SCHA}, the reduction of a generic quadratic Hamiltonian to the normal form introduces the self-consistent harmonic approximation (SCHA). In order to exemplify how the method applies, in Section~\ref{s.spinflopbulk} the so-called spin-flop transition in a bulk Heisenberg antiferromagnet (AFM) is studied, reproducing long-known results~\cite{WangC1964,FederP1968}. The more interesting case of application to a nanosystem is considered in Section\ref{s.spinflopfinite}, namely an antiferromagnetic chain with an {\em odd} number of spins, which displays a peculiar kind of antiferromagnetic phase for low magnetic field, arising from the Zeeman energy gain due to one unbalanced spin, and a spin-flop phase at higher field: establishing the robustness of the ordered phase against quantum fluctuations is a nontrivial achievement. A few closing remarks are eventually drawn in Section~\ref{s.conclusions}.

\section{Weyl symbols}
\label{s.normal-weyl}

Consider the position and momentum operators $\hq$ and $\hp$, with $[\hq,\hp]=i\hbar$. In the following $\hbar$ is not explicitly written, except when significant. Every operator $\hat{O}$ acting on the corresponding Hilbert space is a function of $\hp$ and $\hq$ which can be expressed as
\begin{equation}
 \hat{O}=\int \frac{dr\,dk}{2\pi}~\tilde O(r,k)~e^{i(r\hp+k\hq)} ~;
\label{e.Ork}
\end{equation}
then, the Weyl symbol $O(p,q)$ for $\hat{O}$ is univocally defined~\cite{Weyl1927,AgarwalW1970a,AgarwalW1970b,Berezin1980,Lee1995} as the $c$-number function in phase-space
\begin{equation}
 O(p,q)=\int \frac{dr\,dk}{2\pi}~\tilde O(r,k)~e^{i(rp+kq)} ~.
\label{e.Opq}
\end{equation}
The Fourier transform nature of this definition ensures that it has an inverse, so that there is a one-to-one correspondence between operators and phase-space functions. In particular, the Weyl symbols for $\hp$ and $\hq$ are $p$ and $q$. From the Baker-Hausdoff formula one has $e^{i(r\hp+k\hq)}=e^{ik\hq/2}e^{ir\hp}e^{ik\hq/2}$, which allows one to find the well-known definition of the Weyl symbol in terms of matrix elements,
\begin{equation}
 O(p,q)=\int dr\,\big\langle q{+}{\textstyle\frac{r}2}\big|\,\hat{O}\,
 \big|q{-}{\textstyle\frac{r}2}\big\rangle\,e^{irp} ~.
\end{equation}
Weyl symbols for Hermitian operators are real and their most popular property is the trace formula,
\begin{equation}
 {\rm{Tr}}(\hat{O}\hat{\rho})=\int\frac{dp\,dq}{2\pi}\,O(p,q)\,\rho(p,q) ~.
\label{e.aveO}
\end{equation}
If the generic operator $\hat\rho$ is a density matrix, e.g., at thermal equilibrium, then $(2\pi)^{-1}\rho(p,q)$ is the Wigner distribution function and Eq.~\eqref{e.aveO} is the classical expression for the expectation value of an observable. For instance, if $\hat\rho=|x\rangle\langle{x}|$ then $\rho(p,q)=\delta(q{-}x)$ and the expectation value of $\hat{O}$ is $(2\pi)^{-1}\int{dp}\,O(p,x)=\langle{x}|\hat{O}|{x}\rangle$.
The trace formula does not generalize to the product of three or more operators; the Weyl symbol for the product of two operators is the product of their Weyl symbols only if they commute, otherwise one has to use the `star product'~\cite{Moyal1949,Groenewold1946}.

\subsection{Weyl symbols from normal symbols}

Several physical Hamiltonians are given in normally ordered form, i.e., with all creation operators on the left side: their normal symbols are simply obtained by replacing $\ad$ and $\aa$ with the holomorphic variables $a^*=(q{-}ip)/\sqrt2$ and $a=(q{+}ip)/\sqrt2$. In other cases one deals with operators depending on the number operator $\hat{n}\,{=}\,\ad\aa$, say $\hat{O}=O(\ad\aa)$, whose normal symbol can be written as~\cite{Louisell1973}
\begin{equation}
 O_{_{\rm{N}}}=e^{-a^*a}\sum_{n=0}^\infty O(n)\,\frac{(a^*a)^n}{n!} ~.
\label{e.On}
\end{equation}
That's why explicit relations between Weyl and normal symbols are generally very useful. The definition of the Weyl symbol for an operator $\hat{O}$ through Eqs.~\eqref{e.Ork} and \eqref{e.Opq} has an equivalent in terms of the holomorphic variables, $a=(q{+}ip)/\sqrt2$ and $a^*=(q{-}ip)/\sqrt2$,
\begin{eqnarray}
 \hat{O} = \int \frac{ds\,ds^*}{2\pi i}~\tilde O(s^*,s)~e^{i(s\ad+s^*\aa)} ~,
\\
 O(a^*,a) = \int \frac{ds\,ds^*}{2\pi i}~\tilde O(s^*,s)~e^{i(sa^*+s^*a)} ~;
\label{e.Oaa}
\end{eqnarray}
this can be compared with the `normal symbol' $O_{\rm{N}}(a,a^*)$ for $\hat{O}$, obtained when all creation operators appear on the left side,
\begin{eqnarray}
 \hat{O} = \int \frac{ds\,ds^*}{2\pi i}~\tilde O_{_{\rm{N}}}(s^*,s)~e^{is\ad}\,e^{is^*\aa} ~,
\\
 O_{_{\rm{N}}}(a^*,a) = \int \frac{ds\,ds^*}{2\pi i}
   ~\tilde O_{_{\rm{N}}}(s^*,s)~e^{i(sa^*+s^*a)} ~;
\label{e.Onaa}
\end{eqnarray}
indeed, since $e^{i(s\ad+s^*\aa)}=e^{-\frac12s^*s}\,e^{is\ad}\,e^{is^*\aa}$, it follows that
\begin{equation}
 \tilde{O}(s^*,s)=e^{\frac12s^*s}\,\tilde O_{_{\rm{N}}}(s^*,s) ~;
\label{e.tO-tOn}
\end{equation}
inserting this in Eq.~\eqref{e.Oaa} one can write a formal identity
\begin{equation}
 O(a^*,a) = e^{-\frac12\partial_{a^*}\partial_{a}}\,O_{_{\rm{N}}}(a^*,a) ~.
\label{e.O-On1}
\end{equation}
Let us introduce a Gaussian distribution
\begin{equation}
 P_\alpha(\xi^*,\xi)=\frac1{\pi{i}\alpha}~e^{-\frac{2\xi^*\xi}{\alpha}}
\end{equation}
for the holomorphic variables $(\xi^*,\xi)$; it is fully defined by the averages $\dave{\xi}_\alpha=\dave{\xi^*}_\alpha=0$, $\dave{\xi^2}_\alpha=\dave{(\xi^*)^2}_\alpha=0$ and $\dave{\xi^*\xi}_\alpha=\alpha/2$. Using the identity
\begin{equation}
 \bigdave{e^{i(s\xi^*{\pm}s^*\xi})}_1 
 = e^{-\frac12\dave{(s\xi^*{\pm}s^*\xi)^2}_1}
 = e^{\mp\frac12 s^*s} ~
\end{equation}
to express the relation~\eqref{e.tO-tOn} inside~\eqref{e.Oaa} and~\eqref{e.Onaa}, one finds other integral expressions relating the symbols:
\begin{eqnarray}
 O(a^*,a)&=&\bigdave{O_{_{\rm{N}}}(a^*{+}\xi^*,a{-}\xi)}_1 ~,
 \label{e.O-On2}
\\
 O_{_{\rm{N}}}(a^*,a)&=&\bigdave{O(a^*{+}\xi^*,a{+}\xi)}_1 ~.
\end{eqnarray}
The identities~\eqref{e.O-On1} and~\eqref{e.O-On2} are very useful for obtaining Weyl symbols: let us see two examples.

\subsection{Harmonic oscillator}

For $\hat{H}=\frac\omega{2}\,(\hp^2{+}\hq^2)=\omega\,(\ad\aa{+}\frac12)$, it is $H_{_{\rm{N}}}=\omega\,(a^*a{+}\frac12)$ and Eq.~\eqref{e.O-On1} yields the Weyl symbol $H=\omega\,a^*a=\frac\omega{2}(p^2{+}q^2)$; this is trivial in terms of phase-space symbols. Let us also derive the Weyl symbol for the (non-normalized)  density matrix $\hat\rho_{_\beta}=e^{-\beta\hat H}$ at the equilibrium temperature $\beta^{-1}$. Using Eq.~\eqref{e.On} one finds the normal symbol $\rho_{_{\beta,\rm{N}}}=e^{-f-\kappa\, a^*a}$,
where $f=\beta\omega/2$ and $\kappa=1{-}e^{-\beta\omega}$; then use Eq.~\eqref{e.O-On2}:
\begin{eqnarray}
 \rho_{_\beta} &=& \bigdave{e^{-f-\kappa(a^*{+}\xi^*)(a{-}\xi)}}_1
\notag\\
 &=& e^{-f-\kappa a^*a}{\textstyle \int \frac{d\xi^*d\xi}{\pi{i}}}
 ~e^{-(2{-}\kappa)\xi^*\xi}~e^{\kappa(\xi a^*{-}\xi^*a)}
\notag\\
 &=& e^{-f-\kappa a^*a}\alpha\bigdave{e^{\kappa(\xi a^*{-}\xi^*a)}}_\alpha
\notag\\
 &=& e^{-f-\kappa a^*a}\alpha~e^{-\frac12\kappa^2\alpha~a^*a} ~,
\end{eqnarray}
where $\alpha=\frac2{2{-}\kappa}=\frac{e^f}{\cosh{f}}$. It is then easy matter to get $\kappa+\frac{\kappa^2\alpha}2=2\,\tanh{f}$ and the final result
\begin{equation}
 \rho_{_\beta} = \frac{e^{-2\,a^*a\,\tanh{f}}}{\cosh{f}}
               = \frac{e^{-(p^2{+}q^2)\tanh{f}}}{\cosh{f}} ~;
\end{equation}
the partition function ${\cal Z}=(2\sinh{f})^{-1}$ correctly follows.

\subsection{Operators of the form ~$\hat{F}=f(\ad\aa)\,\aa$}

For an operator $\hat{F}=f(\ad\aa)\,\aa$, taking the Fourier expansion $f(n)=\sum_kc_ke^{ikn}$ and applying Eq.~\eqref{e.On} gives
\begin{equation}
 F_{_{\rm{N}}}=\sum_k c_k\,e^{-(1-e^{ik})a^*a}a~;
\end{equation}
using Eq.~\eqref{e.O-On2} and setting $2\varepsilon\equiv1{-}e^{ik}$
\begin{eqnarray}
 F &=& \sum_k c_k\,\bigdave{e^{-2\varepsilon(a^*{+}\xi^*)(a{-}\xi)}(a{-}\xi)}_1
\notag\\
 &=& \sum_k \frac{c_k}{1{-}\varepsilon}\, e^{-2\varepsilon a^*a}
     \frac1{2\varepsilon}(-\partial_{a^*})
     \bigdave{e^{-2\varepsilon(\xi^*a{-}\xi a^*)}}_{(1{-}\varepsilon)^{-1}}
\notag\\
&=& \sum_k \frac{c_k}{1{-}\varepsilon}\, e^{-2\varepsilon a^*a}
    \frac1{2\varepsilon}(-\partial_{a^*})
    e^{-2\varepsilon^2(1{-}\varepsilon)^{-1}a^*a} 
\notag\\
 &=& \sum_k \frac{c_k\,e^{-ik}}{\cos^2\!\frac{k}2}\, e^{2i\tan\frac{k}2\,a^*a} a ~.
\label{e.Fna1}
\end{eqnarray}
When $f(n)$ is smooth, the main contribution comes from small values of $k$, so
\begin{equation}
 F \simeq \sum_k c_k \, e^{ik\,(a^*a{-}1)} a =f(a^*a{-}1)\,a ~.
\label{e.Fna}
\end{equation}
The above derivation is easily generalized to more general operators, for instance if $\hat{F}_m=f(\ad\aa)\,\aa^m$ then
\begin{equation}
 F_m \simeq f\big(a^*a{-}{\textstyle\frac{1{+}m}2}\big)~a^m~.
\end{equation}

\section{SCHA with Weyl symbols}
\label{s.SCHA}

\subsection{Quadratic Hamiltonians with harmonic normal form}
\label{ss.diagonalH}

As remarked in the Introduction, for a linear system with $n$ degrees of freedom it is easier to perform canonical phase-space transformations, belonging to the symplectic group $Sp(2n,\mathbb{R})$, rather than Bogoliubov transformations~\cite{Colpa1978,Tsallis1978} on the bosonic operators, which belong to the pseudo-unitary group $U(n,n)$. Of course, the two kinds of transformations are in one-to-one correspondence, as proven in Ref.~\onlinecite{Broadbridge1979}. While the classification of the possible normal forms of quadratic Hamiltonians is a well-established subject~\cite{Arnold1978,Broadbridge1979}, the aim here is at giving a simplified approach for dealing with physical Hamiltonians, which have properties such as being bounded from below and being smooth in the neighborhood of their minima, where a quadratic expansion fits the local energy landscape.

Therefore, in this Section an operative procedure is introduced for deriving the linear canonical transformation that reduces to a typical normal form a sufficiently general quadratic Hamiltonian for $N$ degrees of freedom. Without loss of generality, one can work in terms of Weyl symbols and deal with the classical quadratic form
\begin{equation}
 \H_{\rm{q}}(\bp,\bq) = \frac12~\big( \bp^t \bA^2 \bp
 + 2\bp^t\bX\bq + \bq^t\bB^2\bq \big) ~,
\label{e.H2a}
\end{equation}
where the vector notation, $\bp\equiv\{p_i\}$ and $\bq\equiv\{q_i\}$, is used, with $\bA^2$ and $\bB^2$ symmetric real positive-definite $N{\times}N$ matrices, and $\bX$ a real $N{\times}N$ matrix. This Hamiltonian has to have a lower bound and its normal form can be obtained in a few steps.
First, one calculates the positive square-root $\bA$ of $\bA^2$; operatively, one diagonalizes $\bA^2$ with an orthogonal matrix $\bV$, i.e., 
\begin{equation}
 \bV \bA^2 \bV^t=\bLm^2 ~,
\label{e.diagA2}
\end{equation}
where the diagonal matrix $\bLm\equiv{\rm diag}\{\lambda_\ell\}$ contains the positive square-roots of the eigenvalues of $\bA^2$; hence, $\bA=\bV^t\bLm\bV$ and $\bA^{-1}=\bV^t\bLm^{-1}\bV$. The transformation
\begin{eqnarray}
 \bar\bp &=& \bA \bp+\bA^{-1}\bX\bq
\notag\\
 \bar\bq &=& \bA^{-1} \bq~,
\end{eqnarray}
is canonical provided that the matrix $\bX\bA^2$ be symmetric: this is a constraint to be required in order to reduce the Hamiltonian to a sum of harmonic oscillators.~\cite{footnote1} Alternatively, one can exchange the role of coordinates and momenta and require (or check) the symmetry of $\bB^2\!\bX$. The Hamiltonian takes the form $\H=\frac12~\big[\bar\bp^t\bar\bp +\bar\bq^t\bA(\bB^2\!{-}\bX^t\!\bA^{-2}\!\bX)\bA\bar\bq \big]$.
We have then to diagonalize the symmetric real matrix $\bA(\bB^2\!{-}\bX^t\!\bA^{-2}\!\bX)\bA=\bA\bB^2\bA-\bA^{-1}\bX^2\bA$ with an orthogonal $\bO$,
\begin{equation}
 \bO \,\bA(\bB^2\!{-}\bX^t\!\bA^{-2}\!\bX)\bA\, \bO^t
   =  \bOm^2 \equiv{\rm diag}\{\omega_k^2\}.
\label{e.diagAB2A}
\end{equation}
Eventually, performing the canonical transformation
\begin{eqnarray}
 \tilde\bp &=& \bO\bar\bp = \bO\bA\,\bp+\bO\bA^{-1}\bX\bq
\notag\\
 \tilde\bq &=& \bO\bar\bq = \bO\bA^{-1}\bq ~,
\label{e.tildepq}
\end{eqnarray}
the Hamiltonian is expressed in terms of independent harmonic-oscillator modes,
\begin{equation}
 \H_{\rm{q}} = \frac12~\big( \tilde\bp^t\tilde\bp
 + \tilde\bq^t\bOm^2\tilde\bq \big)
 =\frac12\sum_k \big( \tilde p_k^2 + \omega^2_k\,\tilde q_k^2 \big)~.
\end{equation}
The linearity of the transformation~\eqref{e.tildepq} ensures that it also holds for the corresponding operators, $\hat{\tilde\bp}$ and $\hat{\tilde\bq}$, whose Weyl symbols are just $\tilde\bp$ and $\tilde\bq$.
The inverse of Eq.~\eqref{e.tildepq} is
\begin{eqnarray}
 \bp &=& \bF\,\tilde\bp-\bX^t \bF\,\tilde\bq
\notag\\
 \bq &=& \bG\,\tilde\bq ~,
\label{e.bogtr}
\end{eqnarray}
where for shortness of notation the new matrices $\bF$ and $\bG$ are defined such that $\bF^{-1}\,{=}\,\bG^t\,{\equiv}\,\bO\bA$.
It is now easy to express the correlators of the original variables $\bp$ and $\bq$ in terms of the simple correlators of $\tilde\bp$ and $\tilde\bq$. Indeed, assuming the system to be in equilibrium at the temperature $T\equiv\beta^{-1}$, one has
\begin{eqnarray}
 \ave{\tilde\bp\tilde\bp^t}_{kk'}= \ave{\tilde p_k \tilde p_{k'}} &=&\omega_k~\alpha_k~\delta_{kk'}
\notag\\
 \ave{\tilde\bq\tilde\bq^t}_{kk'}= \ave{\tilde q_k \tilde q_{k'}} &=&\omega_k^{-1}\alpha_k~\delta_{kk'}
\notag\\
 \ave{\tilde\bp\tilde\bq^t}_{kk'}= \ave{\tilde p_k \tilde q_{k'}} &=&0 ~,
\label{e.avetildepq}
\end{eqnarray}
where
\begin{equation}
 \alpha_k \equiv \frac{\hbar}{2}~\coth\frac{\beta\hbar\omega_k}{2}~,
\label{e.alphak}
\end{equation}
the frequencies $\omega_k$ being the positive square roots of the eigenvalues $\omega_k^2$; in matrix notation $\bOm=\sqrt{\bOm^2}={\rm{diag}}\{\omega_k\}$. The same equalities hold for the operators corresponding to the Weyl symbols.
In the classical limit $\hbar\,{\to}\,0$ one recovers the equipartition theorem, $\ave{\tilde{p}_k^2}\,{=}\,\omega^2_k\ave{\tilde{q}_k^2}\,{\to}\,T$. Conversely, in the ground state ($T\,{\to}\,0$) one has $\ave{\tilde{p}_k^2}\,{=}\,\omega^2_k\,\ave{\tilde{q}_k^2}
\,{\to}\,\hbar\omega_k/2$, i.e., the zero-point fluctuations of a harmonic oscillator.

For the original momenta and coordinates, the inverse transformation gives the correlators
\begin{eqnarray}
 \ave{\bp\bp^t} &=& \bF\,\ave{\tilde\bp\tilde\bp^t}\,\bF^t
  		+\bX^t\bF\ave{\tilde\bq\tilde\bq^t}\bF^t\bX
\notag\\
 \ave{\bq\bq^t} &=& \bG\,\ave{\tilde\bq\,\tilde\bq^t}\,\bG^t
\notag\\
 \ave{\bp\bq^t} &=& -\bX^t\bF\ave{\tilde\bq\,\tilde\bq^t}\,\bG^t ~,
\end{eqnarray}
which can be written as a sum over $k$-components, for instance
\begin{equation}
 \ave{q_iq_j} = \sum_{k} \frac{\alpha_k}{\omega_k}~G_{ik}G_{jk} ~.
\label{e.correl3}
\end{equation}
It has to be emphasized that these correlators fully define the density matrix (of the ground- or thermal-equilibrium state) or, equivalently, its Weyl symbol, which is a Gaussian distribution in phase space:
\begin{equation}
 \rho(\bp,\bq) \propto \prod_k \exp\Big(
 -\frac{\omega_k^{-1}\tilde p_k^2+\omega_k\tilde q_k^2}{2\alpha_k}\,\Big) ~.
\end{equation}

This means that all higher-order correlators can be expressed in terms of the above ones.

In the zero-temperature case, $\alpha_k=\hbar/2$, one has the ground-state fluctuations
\begin{eqnarray}
 \ave{\bp\bp^t} &=& \frac\hbar{2}~ \bF\,\bOm\,\bF^t
  + \bX^t\bF\,\bOm^{-1}\bF^t\bX
  ~,
\notag\\
 \ave{\bq\bq^t} &=& \frac\hbar{2}~ \bG\,\bOm^{-1}\bG^t ~.
\label{e.correl4}
\end{eqnarray}
Note that for $\bX=0$, as $\bF^t\bG=\mathds{1}$, one finds the minumum Heisenberg's uncertainty,
\begin{equation}
 \ave{\bp\bp^t}\,\ave{\bq\bq^t} = \frac{\hbar^2}{4}~\mathds{1}~.
\label{e.Heispr}
\end{equation}
The transformations considered in this Section are easily implemented numerically for systems with few degrees of freedom, as they only involve real matrices and diagonalizations of real symmetric matrices. This is a well established starting point for implementing a procedure that accounts for the interaction, i.e., non-quadratic terms in the Hamiltonian.

\subsection{SCHA for a one-dimensional particle}
It is easy to explain what underlies the idea of the SCHA. Take for instance a single classical particle of unit mass in a one-dimensional potential $V(q)$ which has its minimum in $q\,{=}\,0$, say  $V(0)=0$. For nonzero temperature, the exact (non normalized) coordinate distribution is given by $\rho(q)=e^{-\beta\,V(q)}$. For very low temperature it makes sense to approximate $V(q)\simeq\frac12{V''}(0)\,q^2$, yielding a Gaussian distribution: this is the harmonic approximation (HA). At higher temperature thermal fluctuations become larger and it becomes much better to use a `trial' harmonic potential
\begin{equation}
  V_0(q)=w+\frac{\omega^2}2\,(q{-}q_0)^2 ~,
\label{e.V0}
\end{equation}
with the associated Gaussian distribution $\rho_0(q)=e^{-\beta\,V_0(q)}$, such that $\bigave{(q{-}q_0)^2}_0=(\beta\omega^2)^{-1}$, where $\ave{\cdots}_0$ denotes the average with the Gaussian distribution $\rho_0$. The parameters $w$, $q_0$, and $\omega^2$ can be determined variationally, by minimizing the r.h.s. of the Bogoliubov inequality,
\begin{equation}
 F\leq F_0+\bigave{V(q)-V_0(q)}_0 ~.
\end{equation}
This minimization is equivalent to requiring that $V(q)$ and $V_0(q)$ have the same trial averages, together with their first and second derivatives:
\begin{eqnarray}
 \bigave{V(q)}_0&=&w+\frac{\omega^2}2\,\bigave{(q{-}q_0)^2}_0 ~,
\notag\\
 \bigave{V'(q)}_0&=&\omega^2\ave{q{-}q_0}_0=0 ~,
\notag\\
 \bigave{V''(q)}_0&=&\omega^2 ~.
\label{e.schaV}
\end{eqnarray}
In this way $V_0$ approximates $V$ in a way that accounts for the statistical importance of the coordinate; for instance the renormalized square frequency $\omega^2$ accounts for the average curvature of the original potential. It is obvious, since $\ave{\cdots}_0$ depends on $\omega^2$ and $q_0$, that the last two equations are to be solved self-consistently (SC), e.g., by iteration: hence the designation SCHA. The parameters $q_0$, and $\omega^2$, as well as $w$ (that follows from the first condition), depend on $\beta$; $\omega=\omega(\beta)$ is usually dubbed `renormalized' frequency.

The above framework also holds for quantum systems: one has just to identify $\rho_0(q)$ with the quantum coordinate distribution for the harmonic potential~\eqref{e.V0}, which amounts to set $\bigave{(q{-}q_0)^2}_0=\frac{\hbar}{2\omega}\coth\frac{\beta\hbar\omega}2$. Due to ground-state fluctuations, in the quantum case one does not recover the HA at zero temperature: the SCHA accounts for anharmonicity also in the ground state. It may happen that $\omega^2={0}$ for some temperature $\beta_{\rm{c}}$: this signals the impossibility to fit $V(q)$ with the trial potential and can be interpreted as the onset of an instability leading towards a different stable state. In the case of many degrees of freedom, the vanishing of the spectrum gap often corresponds to a phase transition; however, such an instability can also appear as a function of the Hamiltonian's parameters, e.g., a magnetic field, which case is commonly referred to as a `quantum phase transition', even when its basic mechanism is essentially classical, as it happens, e.g., for the Ising chain in a transverse field~\cite{CTVV2007prb,CTVV2007jmmm}, for the Kosterlitz-Thouless transition in the 2D $XXZ$ model~\cite{CTVV1995prb}, and for 2D frustrated antiferromagnets~\cite{CCTVV1999prb}.

The SCHA has had several successful applications in the study of quantum many-body systems, initially in lattice dynamics~\cite{Hooton1955,Koehler1966,KleinH1972} (often called `SC phonon theory'), but afterwards in optics, in magnetism (`SC spin-wave theory'), and many other fields. A similar approach has also been exploited within Feynman's path integral formalism, and has found an improved extension with plenty of applications in the study of quantum systems~\cite{GT1985,FK1986,CGTVV1995}.

\subsection{SCHA for general Hamiltonians}

Few-body nanostructures are somewhat more subtle than their extended translation-symmetric counterparts: the quantum SCHA is here derived in terms suitable for a simplified (possibly numerical) approach.

For a single degree of freedom, in the general case one has to consider the Weyl symbol for the Hamiltonian, $\H(p,q)$. Then Eqs.~\eqref{e.schaV} are readily generalized to the request that the six parameters of the trial quadratic Hamiltonian,
\begin{equation}
 \H_0(p,q)=w+a(p{-}p_0)^2+b(q{-}q_0)^2+c(p{-}p_0)(q{-}q_0) ~,
\end{equation}
be determined by imposing the equality of the $\ave{\cdots}_0$ averages of $\H$ and $\H_0$, and of their derivatives up to second order, which indeed amounts to six conditions.  

For many degrees of freedom one has to approximate the Hamiltonian $\H(\bp,\bq)$ using a trial Hamiltonian in the general form
\begin{equation}
 \H_0(\bp,\bq)=w+\H_{\rm{q}}(\bp{-}\bp_0,\bq{-}\bq_0)
\label{e.H0}
\end{equation}
where $\H_{\rm{q}}$ is the quadratic form~\eqref{e.H2a}. The parameters to be determined are $w$, the $N$-component vectors $\bp_0$ and $\bq_0$, and the $N{\times}N$-matrices $\bA^2$, $\bB^2$, $\bX$. Extending the above conditions to the $\ave{\cdots}_0$ averages one can in principle determine all the parameters. However, in order to find $\rho_0(\bp,\bq)$ one has to perform the (parameter-dependent) canonical transformation that reduces $\H_0$ to the standard normal form, as made in Section~\ref{ss.diagonalH}. This requires $\bX\bA^2$ be symmetric~\cite{footnote1}: fortunately, in the majority of systems one encounters in Matter Physics, interaction terms between momenta and coordinates do not appear and the matrix $\bX$ vanishes. A partial exception are spin Hamiltonians in the bosonic representation, where such terms do appear, but the corresponding SCHA entails $\bX\,{=}\,0$.

An important tool when calculating the Gaussian averages $\ave{\cdots}_0$ is the known technique of `Wick decoupling', that follows from a property of Gaussian distributions: the Gaussian average of a nonlinear $2n$-degree monomial in the stochastic variables (assumed with vanishing average) can be obtained by separating it in all possible ways (considering all variables as different) as product of quadratic couples, replacing the couples with the corresponding average, and eventually summing up all the $(2n{-}1)!!$ terms. For instance (when $\ave{p}_0=\ave{q}_0=0$) one has $\ave{q^{2n}}_0=(2n{-}1)!!~\ave{q^2}_0^n$ and $\ave{p^2q^4}_0=3\ave{p^2}_0\ave{q^2}_0^2+12\ave{pq}_0^2\ave{q^2}_0$. Of course odd-degree terms average to zero.

The next Sections show how the SCHA with Weyl symbols applies to the study of two different kinds of transition in Heisenberg AFMs, both with and without translation symmetry.

\section{Spin-flop field in the bulk AFM}
\label{s.spinflopbulk}

\subsection{Easy-axis AFM in a field}
\label{ss.eaAFM}

\begin{figure}
	\includegraphics[width=0.45\textwidth]{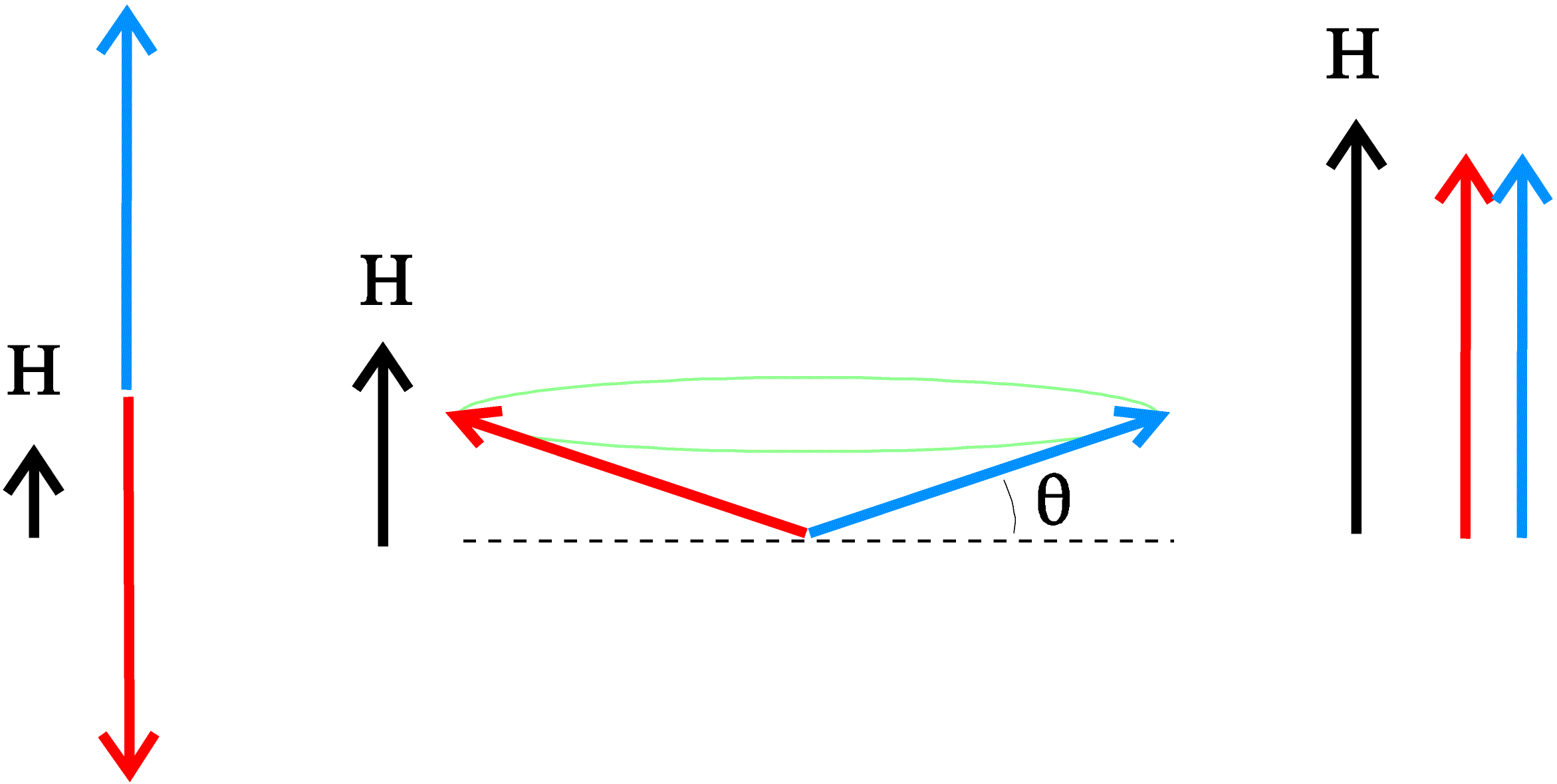}
	\caption{Minimum energy configuration of (the classical counterpart of) the AFM described by~\eqref{e.Hafm}: for small field, the spins of the two sublattices are antiparallel along the $z$-axis; beyond a given field $H_{_{\rm{SF}}}$ the spin-flop configuration prevails, the angle $\Theta$ increasing with $H$ up to $\pi/2$ at the saturation field.}
	\label{f.spinflop}
\end{figure}

Let us consider an AFM on a bipartite lattice, with {\em easy-axis} exchange anisotropy, $\mu\gtrsim1$, in a magnetic field $\bm{H}$ parallel to the easy axis. The Hamiltonian reads
\begin{equation}
 \hat\H=\frac{J}2\sum_{\bi,\bd}\big( 
 \hat S_\bi^x\hat S_{\bi+\bd}^x + \hat S_\bi^y\hat S_{\bi+\bd}^y
 +\mu\hat{S}^z_\bi \hat{S}^z_{\bi+\bd}\big)  -H\sum_\bi \hat{S}^z_\bi~.
\label{e.Hafm}
\end{equation}
$\hat{\bm{S}}_\bi$ are spin operators belonging to the spin-$S$ representation, and the index $\bi$ runs over all $N$ lattice sites, while $\bd$ connects any site $\bi$ with its $z$ nearest-neighbors at $\bi{+}\bd$, all belonging to the other sublattice; periodic boundary conditions are assumed. The ground state for vanishing or sufficiently small field is close to the classical {\em N\'eel} state, with neighboring spins antiparallel (AP) along the $z$-direction. The goal is to determine the `critical' field value $H_{_{\rm{SF}}}$ where this state becomes unstable (in favor of the spin-flop state). The problem is easily solved in the classical translation-invariant case, as shown in Fig.~\ref{f.spinflop}: the classical spins are vectors of length, say, $\tS$ and the energy of the spin-flop configuration, occurring for  $\sin\theta=H/[J\tS{z}(1{+}\mu)]$, is smaller than that of the AP one if $H>H_{_{\rm{SF}}}^{(\rm{cl)}}$, where
\begin{equation}
 H_{_{\rm{SF}}}^{(\rm{cl)}}=J\tS{z}\sqrt{\mu^2{-}1} ~.
\label{e.Hsfcl}
\end{equation}

Quantum spins are conveniently treated by means of the Holstein-Primakoff (HP) transformation~\cite{HolsteinP1940}, which transforms spin operators to bosonic operators preserving the spin commutation relations,
\begin{eqnarray}
\hat S^+&=&\hat S^x+i\hat S^y=(2S{-}\ad\aa)^{1/2}~\aa
\notag\\
\hat S^z &=& S-\ad\aa ~,
\label{e.HP}
\end{eqnarray}
and in this form it is suitable for describing states where $\hat{S}^z$ is close to $S$; indeed, in spite of the constraint that the boson number cannot exceed $2S$, this is a very good approach~\cite{Dyson1956,Oguchi1960} when interactions and temperatures do not make $S^z$ to deviate too much from $S$.

In order to transform spins in the two sublattices, two families of bosonic operators are usually introduced~\cite{Oguchi1960}, but this complication can be overcome by turning the spins belonging to the `down' sublattice by the canonical transformation
\begin{equation}
 \hat{\bm{S}} = (\hat S^x,\hat S^y,\hat S^z) \longrightarrow
 (-\hat S^x,\hat S^y,-\hat S^z)~,
\label{e.canonical}
\end{equation}
which makes the Hamiltonian~\eqref{e.Hafm} `ferromagnetic'.
\begin{eqnarray}
 \hat\H &=& -\frac{J}2\sum_{\bi,\bd}\Big[ {\textstyle\frac12}
  \big( \hat S_\bi^+\hat S_{\bi+\bd}^+ + \hat S_\bi^-\hat S_{\bi+\bd}^-\big) 
  +\mu\,\hat S_\bi^z\hat S_{\bi+\bd}^z\Big]
\notag\\
 && + H \sum_{\bi} (-)^\bi\hat S_i^z~;
\label{e.Hferro}
\end{eqnarray}
$(-)^\bi$ is the `sign' of the sublattice to which the site $\bi$ belongs. With the new variables the magnetic field becomes {\em staggered} and the AP (N\'eel) configuration is converted into the ferromagnetic one, which is not an eigenstate of $\hat\H$ due to the terms $\hat S_\bi^-\hat S_{\bi+\bd}^-$.

The Weyl symbol for the longitudinal spin component is immediately found by Eq.~\eqref{e.O-On1}, since $S^z_{_{\rm{N}}}=S-a^*a$,
\begin{equation}
 S^z = \tS-a^*a~,
\end{equation}
where
\begin{equation}
 \tS \equiv S+{\textstyle\frac12} ~;
\label{e.stilde}
\end{equation}
for the transverse components Eq.~\eqref{e.Fna} gives the Weyl symbol
\begin{equation}
 S^+ \simeq (2\tS{-}a^*a)^{1/2}a ~;
\end{equation}
note that the above mapping corresponds to a classical spin of length $|{\bm{S}}|=\tS$. Turning to phase-space variables $(p,q)$, i.e., the Weyl symbols for the momentum and coordinate operators,
\begin{equation}
 \hat p=i\frac{\hat a^\dagger{-}\hat a}{\sqrt{2\tS}}
~,~~~~~~~~~
 \hat q=\frac{\hat a^\dagger{+}\hat a}{\sqrt{2\tS}} ~,
\label{e.hatpq}
\end{equation}
and setting
\begin{equation}
  r\equiv\frac{q^2{+}p^2}4 ~,
\label{e.r}
\end{equation}
the Weyl symbol for the spin operator $\hat{\bm{S}}$ reads~\cite{CTVV1997}
\begin{equation}
 {\bm{S}} =\tS \Big(\sqrt{1{-}r}~q,\sqrt{1{-}r}~p,1{-}2r\Big) ~.
\label{e.WeylHP}
\end{equation}
The `Weyl spin length' $\tS$, Eq.~\eqref{e.stilde}, plays a relevant role since its reciprocal replaces Planck's constant, $[\hat q,\hat p]=i\,\tS^{-1}$, consistently with the fact that the classical limit coincides with $\tS\,{\to}\,{\infty}$. The Weyl symbol for the Hamiltonian~\eqref{e.Hferro} is obtained by replacing the individual spin operators with their Weyl symbols,
\begin{eqnarray}
 &&\frac\H{J\tS^2} = -\frac12 \sum_{\bi,\bd}\bigg[
 \sqrt{1{-}r_\bi} \sqrt{1{-}r_{\bi+\bd}}
 (q_\bi q_{\bi+\bd}{-}p_\bi p_{\bi+\bd})
\notag\\
 &&~
 +\mu(1{-}2r_\bi)(1{-}2r_{\bi+\bd}) \bigg]
 + h \sum_\bi (-)^\bi (1{-}2r_\bi) ~.
\label{e.HWafm}
\end{eqnarray}
Here
\begin{equation}
 h\equiv\frac{H}{J\tS}
\end{equation}
is the reduced field and from now on $J\tS^2$ is taken as the overall energy scale (i.e., $J\tS^2\,{=}\,1$). This Hamiltonian displays a remarkable symmetry: the (canonical) transformation of changing the signs of both $p_\bi$ and $q_\bi$ in one sublattice exchanges the roles of momenta and coordinates. This is connected to the fact that in the transformation~\eqref{e.canonical} the choice of changing the sign of $\hat{S}^y$ rather than that of $\hat{S}^x$ is arbitrary. As a consequence one can immediately establish some equalities, such as
\begin{equation}
 \ave{p_\bi p_\bi}=\ave{q_\bi q_\bi}
~,~~~~~~~
 \ave{p_\bi p_{\bi+\bd}}=-\ave{q_\bi q_{\bi+\bd}} ~,
\label{e.corrsymm}
\end{equation}
where the average is taken in the ground state or in a thermal equilibrium state. The Hamiltonian~\eqref{e.HWafm} contains only even-order terms,
\begin{eqnarray}
 \H &=& -{\textstyle\frac12} {Nz}\mu + \H^{(2)} + \H^{(4)} + \cdots
\\
 \H^{(2)} &=& \frac12 \sum_{\bi,\bd}\big[
 p_\bi p_{\bi+\bd}{-}q_\bi q_{\bi+\bd}+4\mu\, r_\bi \big]
\notag\\ &&~
 - 2h \sum_\bi (-)^\bi r_\bi ~,
\label{e.H2}
\\
 \H^{(4)} &=& \sum_{\bi,\bd} r_\bi \big[
 {\textstyle\frac12} (q_\bi q_{\bi+\bd}{-}p_\bi p_{\bi+\bd})
 -2\mu\, r_{\bi+\bd} \big] ~.
\label{e.H4}
\end{eqnarray}
The leading quantum effects are to be studied by the SCHA including the interaction $\H^{(4)}$.

\subsection{Harmonic approximation}
\label{ss.AFM.HA}

Translation symmetry entails that the degrees of freedom in Fourier space become uncoupled. Indeed, by defining
\begin{equation}
 q_\bi=\frac1{\sqrt{N}}\sum_\bk e^{i \bk{\cdot}\bi}q_\bk
 ~,~~~~~
 p_\bi=\frac1{\sqrt{N}}\sum_\bk e^{i \bk{\cdot}\bi}p_\bk
\label{e.FTpq}
\end{equation}
one finds
\begin{eqnarray}
 \H^{(2)} &=& \frac12\sum_\bk\big[z(\mu{+}\gamma_\bk)p^2_\bk
                               +z(\mu{-}\gamma_\bk)q^2_\bk
\notag\\
 &&~~~~~~~~ - h (p_\bk p_{\bk+\bpi}+q_\bk q_{\bk+\bpi})\big] ~,
\label{e.H2k.1}
\end{eqnarray}
where $\bpi$ is such that $e^{i\bpi{\cdot}\bi}=(-)^\bi$ and
\begin{equation}
 \gamma_\bk\equiv\frac1z \sum_\bd e^{i\bk{\cdot}\bd}=-\gamma_{\bk+\bpi}~,
\end{equation}
the last equality arising because the sites $\bi$ and $\bi{+}\bd$ belong to opposite sublattices, so $e^{i\bpi{\cdot}(\bi{+}\bd)}=-e^{i\bpi{\cdot}\bi}$, or $e^{i\bpi{\cdot}\bd}=-1$. 

The above quadratic form is diagonal when $h\,{=}\,0$, the corresponding frequencies being $\omega_\bk=z\sqrt{\mu^2{-}\gamma^2_\bk}$. When $h\,{>}\,0$ it can be diagonalized by summing over half of the Brillouin zone (denoted by the prime)
\begin{eqnarray}
\H^{(2)} &=& \frac12 {\sum_\bk}'\big[z(\mu{+}\gamma_\bk)p^2_\bk+z(\mu{-}\gamma_\bk)q^2_\bk
 +z(\mu{-}\gamma_\bk)p^2_{\bk+\bpi}
\notag\\
 &&+z(\mu{+}\gamma_\bk)q^2_{\bk+\bpi}
 - 2h (p_\bk p_{\bk+\bpi}+q_\bk q_{\bk+\bpi})\big] ~,
\label{e.H2k.2}
\end{eqnarray}
The structure is $\H^{(2)}=z\sum'_\bk\H_\bk$, with
\begin{equation}
 \H_\bk = \frac12\,\big(\bp_\bk^t\bA^2_\bk\bp_\bk+\bq_\bk^t\bB^2_\bk\bq_\bk\big) ~,
\label{e.H2k.3}
\end{equation}
with two-dimensional vectors and matrices
\begin{eqnarray}
 \bq_\bk = \binom{q_\bk}{q_{\bk+\bpi}}
~,&&~~~
 \bB^2_\bk = \begin{pmatrix} \mu{-}\gamma_\bk & -\bar{h} \\ -\bar{h} & \mu{+}\gamma_\bk
 \end{pmatrix}
~,\\
 \bp_\bk = \binom{p_\bk}{p_{\bk+\bpi}}
 ~,&&~~~
 \bA^2_\bk = \begin{pmatrix} \mu{+}\gamma_\bk & -\bar{h} \\ -\bar{h} & \mu{-}\gamma_\bk \end{pmatrix}~,
\end{eqnarray}
and $\bar{h}\,{=}\,h/z$. The two matrices do not commute, but have the same eigenvalues, $\mu{\pm}(\bar{h}^2{+}\gamma_\bk^2)^{1/2}$, and are positive definite if $\bar{h}^2<\mu^2{-}\gamma_\bk^2$, so that at this level of approximation one has the classical result $h_{_{\rm{SF}}}^{\rm{(cl)}}=z\sqrt{\mu^2{-}1}$.

Consider now the subspace with two coupled degrees of freedom labeled by $\bk$ and $\bk{+}\bpi$ and the  Hamiltonian Eq.~\eqref{e.H2k.3}; for simplicity, subscripts can be understood. The eigenmodes (for $h\,{<}\,h_{_{\rm{SF}}}$) can be found by the method of Section~\ref{ss.diagonalH} (with $\bX\,{=}\,0$), and Eq.~\eqref{e.diagAB2A} reads $\bO\bA\bB^2\bA\bO^t=\bOm^2={\rm{diag}}(\omega^2_\pm)$. From the identity $(\bA^2,\bB^2)=\mu\pm\gamma\sigma_3-\bar{h}\sigma_1$ ($\sigma_\alpha$ are the Pauli matrices) it follows that $\bB^2=\sigma_1\bA^2\sigma_1$, so $\bA\bB^2\bA=(\bA\sigma_1\bA)^2$ and it suffices to diagonalize $\bO(\bA\sigma_1\bA)\bO^t=\tilde\bOm={\rm{diag}}(\tilde\omega_\pm)$ (not positive definite!) to find $\bOm^2=\tilde\bOm^2$. Rewriting the eigenvalue equation as $(\bA^2\sigma_1)\,\bA\bO^t=\bA\bO^t\tilde\bOm$, it is immediate to find $\tilde\omega_\pm=\pm\sqrt{\mu^2{-}\gamma^2}-\bar{h}$: therefore the eigenfrequencies are $\omega_\pm=|\tilde\omega_\pm|=\sqrt{\mu^2{-}\gamma^2}\mp\bar{h}$; one also finds that  
\begin{equation}
 \bA\bO^t=
  \begin{pmatrix} c_+\sqrt{\mu{+}\gamma} & ~c_-\sqrt{\mu{+}\gamma} \\
  	              c_+\sqrt{\mu{-}\gamma} & -c_-\sqrt{\mu{-}\gamma} \end{pmatrix}~,
\end{equation} 
where the normalizations $c_\pm$ can be determined by requiring the orthogonality of $\bO$ in the form $(\bA\bO^t)(\bA\bO^t)^t\,{=}\,\bA^2$,
\begin{equation}
c_\pm^2=\frac{\sqrt{\mu^2{-}\gamma^2}\mp\bar{h}}{2\sqrt{\mu^2{-}\gamma^2}}
       =\frac{\omega_\pm}{\omega_+{+}\omega_-}~.
\end{equation}
From the ground-state quantum fluctuations of the harmonic oscillator, $\ave{p^2}=\omega/(2\tS)$ and $\ave{q^2}=1/(2\tS\omega)$ (remind that $\hbar\leftrightarrow\tS^{-1}$), it follows that the canonical variables have the correlation matrices
\begin{eqnarray}
 2\tS~\bigave{\bq\bq^t} &=& \bA\bO^t\,\bOm^{-1}\,\bO\bA~,
\notag\\
 2\tS~\bigave{\bp\bp^t} &=& \bA^{-1}\bO^t\,\bOm\,\bO\bA^{-1} ~,
\end{eqnarray}
which, by the way, entail the Heisenberg principle $\bigave{\bp\bp^t}\bigave{\bq\bq^t} = (2\tS)^{-2}$.

Using the identity $\bO^t\,\bOm\,\bO=|\bA\sigma_1\bA|$ one can prove that $\bigave{\bp\bp^t}=\sigma_1\bigave{\bq\bq^t}\sigma_1$, so it is sufficient to calculate
\begin{equation}
  2\tS~\bigave{\bq\bq^t} = 
  {\rm{diag}}\Big(\sqrt{\frac{\mu{+}\gamma}{\mu{-}\gamma}},
   \sqrt{\frac{\mu{-}\gamma}{\mu{+}\gamma}}~\Big)~.
\end{equation}
Surprisingly enough, the fluctuations are decoupled (diagonal matrix) and do not depend on $h$.

Restoring the Fourier subscripts,
\begin{equation}
 2\tS~\bigave{q_\bk q_{\bk'}} = \delta_{\bk\bk'}~\sqrt{\frac{\mu{+}\gamma_\bk}{\mu{-}\gamma_\bk}}~.
\label{e.qqAFM}
\end{equation}
Using Eqs.~\eqref{e.FTpq} one can build the correlators
\begin{eqnarray}
 D&\equiv& \ave{q_\bi^2}=\frac1{2\tS N}\sum_\bk \sqrt{\frac{\mu{+}\gamma_\bk}{\mu{-}\gamma_\bk}}
\notag\\
 D'&\equiv& \ave{q_\bi q_{\bi+\bd}}=\frac1{2\tS N}\sum_\bk \gamma_\bk\, \sqrt{\frac{\mu{+}\gamma_\bk}{\mu{-}\gamma_\bk}} ~.
\label{e.DD1AFM}
\end{eqnarray}
The bare eigenfrequencies are
\begin{equation}
 \omega_{\pm,\bk}=z~\sqrt{\mu^2{-}\gamma_\bk^2}~\pm{h}~,
\label{e.omkpm}
\end{equation}
and the instability $\omega_{-,{\bm{0}}}\,{=}\,0$ occurs indeed at the field value $h_{_{\rm{SF}}}^{\rm{(cl)}}\,{=}\,z\sqrt{\mu^2{-}1}$, i.e., Eq.~\eqref{e.Hsfcl}.

Note that already at this level the results agree with the usual bosonic approach: for instance, from Eqs.~\eqref{e.r} and~\eqref{e.WeylHP}, the ground-state staggered magnetization is given by $m\,{=}\,\tS(1{-}D)$. In the isotropic case $\mu\,{=}\,1$ this agrees with the known result, e.g., Eq.~(31) of Ref.~\onlinecite{WangC1964}, obtained using creation/annihilation operators. These have become popular since systematic perturbative approaches were introduced by famous textbooks~\cite{AbrikosovGD1961,FetterW1971} and calculating the effects of quantum fluctuations starting from the bosonic vacuum was considered a good deal. The above treatment shows that one can equally start from the `classical' Weyl-symbol picture to get the very same estimations in a transparent way. The difference between the bosonic and the phase-space approach is well symbolized by the fact that the same correct staggered magnetization $m$ is obtained as a correction to the vacuum value $m\,{=}\,S$ and to the classical value $m\,{=}\,\tS$, respectively.

\subsection{Interaction and SCHA}
\label{ss.AFM.SCHA}

It is not necessary to account for a trial coupling matrix $\bX$, because the Hamiltonian~\eqref{e.HWafm}, and in particular its quartic part~\eqref{e.H4}, does not contain terms of odd order in the $q$'s and the $p$'s, so it is consistent with the vanishing of all correlators $\ave{\H_{p_iq_j}}_0$. As for the averaged first derivatives, they vanish for the same reason since $\ave{p_i}_0=\ave{q_i}_0=0$, i.e., $\bp_0=\bq_0=0$ and the AP configuration is not `shifted' by quantum (and thermal) fluctuations. One then proceeds by taking the second derivatives of Eq.~\eqref{e.H4},
\begin{eqnarray}
 4\, \H^{(4)}_{q_\bi q_\bi} &=& \sum_\bd
 \big[3q_\bi q_{\bi+\bd} - p_\bi p_{\bi+\bd}
 -2\mu (q_{\bi+\bd}^2{+}p_{\bi+\bd}^2) \big]
\nonumber\\
8\, \H^{(4)}_{q_\bi q_{\bi+\bd}} &=&
3q_\bi^2{+}3q_{\bi+\bd}^2+p_\bi^2{+}p_{\bi+\bd}^2 {-}8\mu\, q_\bi q_{\bi+\bd} \,,
\end{eqnarray}
and, using the symmetry relations~\eqref{e.corrsymm},
\begin{eqnarray}
 \ave{\H^{(4)}_{q_\bi q_\bi}}_0
  &=& \ave{\H^{(4)}_{p_\bi p_\bi}}_0
 = z \big(\ave{q_\bi q_{\bi+\bd}}_0-\mu\ave{q_{\bi}^2}_0\big)
\notag\\ &=& z(D'{-}\mu D)
\notag\\
 \ave{\H^{(4)}_{q_\bi q_{\bi+\bd}}}_0
  &=& -\ave{\H^{(4)}_{p_\bi p_{\bi+\bd}}}_0
 = \ave{q_\bi^2}_0-\mu\ave{q_\bi q_{\bi+\bd}}_0
 \notag\\ &=& D{-}\mu D'~.
\label{e.H4symm}
\end{eqnarray}
These determine the 
`dressed' parameters 
\begin{equation}
\theta = 1{-}D{+}\mu D'
~,~~~~~
\tilde\mu = \mu(1{-}D){+}D' ~.
\end{equation}
that appear in the renormalized Hamiltonian~\eqref{e.H0}
\begin{eqnarray}
 \H_0 &=& \frac12 \sum_{\bi,\bd}\big[
 \theta (p_\bi p_{\bi+\bd}{-}q_\bi q_{\bi+\bd})
 +\tilde\mu (p_\bi^2{+}q_\bi^2) \big]
\notag\\
 && \hspace{30mm} - h \sum_\bi (-)^\bi r_\bi ~.
 \label{e.H0AFM}
\end{eqnarray}
This can be interpreted as a renormalization of the matrices $\bA$ and $\bB$, which eventually results in the renormalized frequencies $\tilde\omega_k$. The corrections are larger the smaller the spin value, since they involve averages of quadratic terms that are of order $1/\tS$. 
The self-consistent correlators $D$ and $D'$ follow by replacing $\mu\to\tilde\mu$ and $\gamma_\bk\to\theta\gamma_k$ in Eqs.~\eqref{e.DD1AFM},
while the renormalized spectrum $\omega_{\pm,\bk}=z\sqrt{\tilde\mu^2-\theta^2\gamma_\bk^2}\pm{h}$ entails the quantum effect upon the spin-flop field
\begin{equation}
 h_{_{\rm{SF}}} = z\sqrt{\tilde\mu^2-\theta^2}
             = h_{_{\rm{SF}}}^{\rm{(cl)}}\, (1{-}D) +O(\tS^{-2}) ~,
\label{e.hsf}
\end{equation}
which was the goal of the present example: the transition field becomes smaller than its classical counterpart, meaning that quantum fluctuations make the ordered AP configuration more unstable towards spin-flop. This comes to no surprise, as it is very common to find that quantum fluctuations weaken the stability of ordered states. It is easy to prove that Eq.~\eqref{e.hsf} agrees with (the $T\to{0}$ limit of) Eq.~(32) of Ref.~\onlinecite{FederP1968}, where the easy-axis anisotropy is of the single-site type, the correspondence being $\mu\leftrightarrow\tilde{K}$.

\section{AFM chain with an odd number of spins}
\label{s.spinflopfinite}

Consider now a {\em finite} chain described by Eq.~\eqref{e.Hafm}, i.e., the lattice is one-dimensional and the number $N$ of spins is finite, with {\em open boundary conditions}. For the classical system it has been discovered~\cite{LounisDB2008,PolitiP2009,BerzinMS2010} that when the number of spins is {\em odd}, $N=2M{+}1$, the AP configuration (with $M$ down and $M{+}1$ up spins) can be stable for finite field even if the anisotropy is slightly {\em easy-plane}, $\mu\lesssim{1}$. Such a {\em parity effect} is due to the Zeeman energy gain of one unbalanced spin which wins against the spin-flop configuration up to a transition field $h_+(N,\mu)$. Furthermore, it is intuitive that for easy-plane anisotropy at zero-field the spins lie in the easy-plane and the AP configuration sets in one only beyond a finite field $h_-(N,\mu)$: hence, by increasing the field there are two boundaries enclosing the AP phase, i.e., a re-entrant transition~\cite{CPV2014,footnote2}, as shown in Fig.~\ref{f.phased}. Remarkably, this classical phase boundary can be calculated analytically, as sketched in Appendix~\ref{a.ha}, in spite of the complexity of the spin-flop configuration~\cite{TralloriPRMP1996,PolitiP2009} that takes place beyond it. For large $N$ this phenomenology is washed out, so the boundaries $h_\pm(N,\mu)$ have to collapse beyond some `critical' value of $N$. Therefore, in these short chains the AP phase seems to be fragile, and a question arises: Does this classical picture also hold for quantum spins or is the AP state washed out by quantum fluctuations? The SCHA can give an answer.

\begin{figure}
	\includegraphics[width=0.45\textwidth]{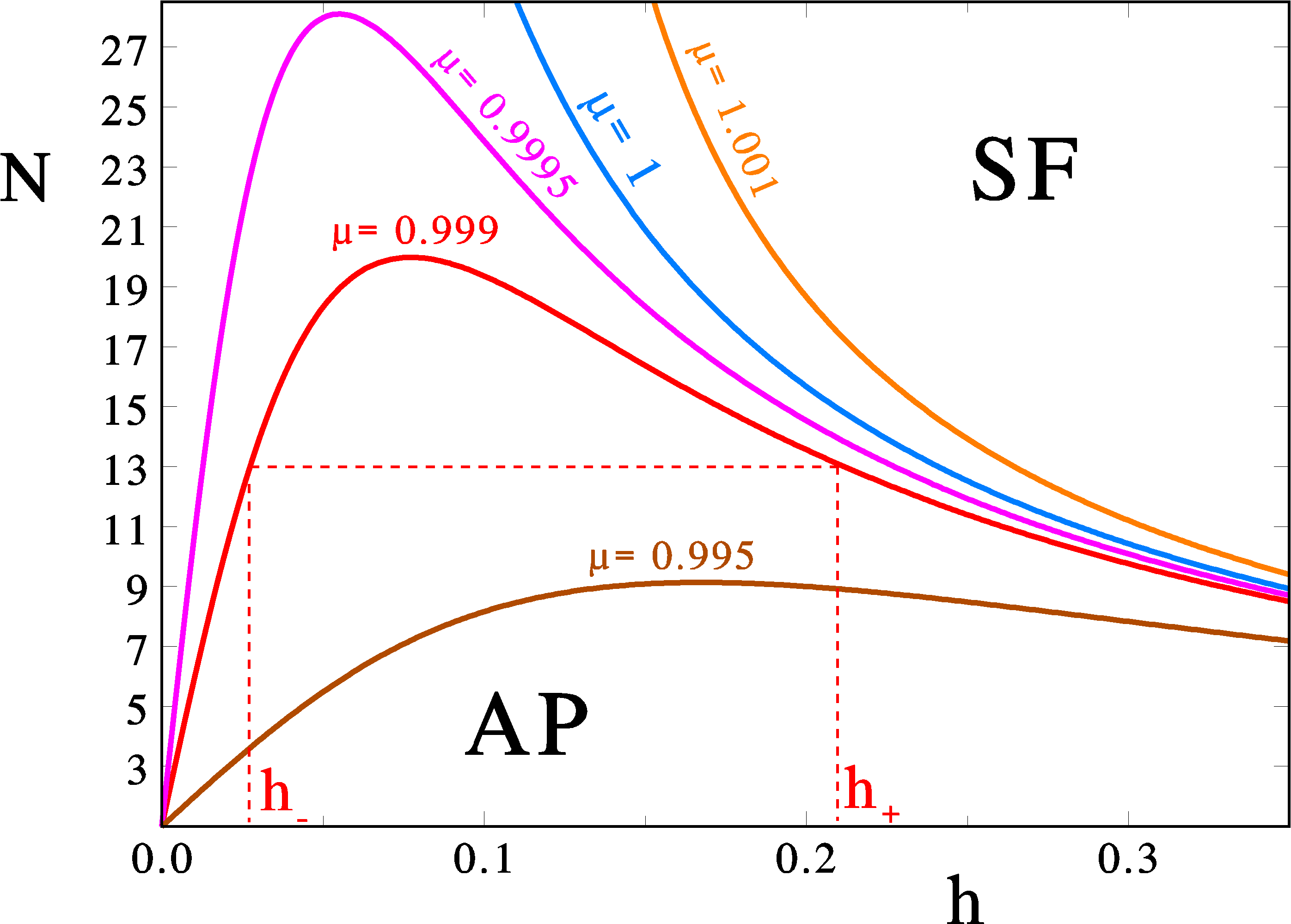}
	\caption{Phase diagram for the finite chain with an odd number of classical spins $N$. The curves report the analytical result derived in Appendix~\ref{a.ha}, which interpolates the points at odd values of $N$. For easy-plane anisotropy $\mu\,{<}\,1$ the AP configuration exists between two field values $h_\mp$ (those for $\mu\,{=}\,0.999$ and $N\,{=}\,13$ are shown), and there is a maximum $N$ beyond which only the SF phase exists; for $\mu\,{\ge}\,1$ the AP phase persists for $h\,{\to}\,0$, i.e., $h_-\,{=}\,0$, and the AP phase exist foe any $N$.}
	\label{f.phased}
\end{figure}

Note that the Hamiltonian is still symmetric for rotations around the $z$-axis, i.e., those generated by the total $z$-component of the magnetization, $\hat{M}^z=\sum_i\hat{S}_i^z$, which indeed commutes with $\hat\H$. Therefore, its ground state has a definite value of $\hat{M}^z\,{=}\,M^z$. For sufficiently small $h$, the magnetization has to be the lowest possible value, namely $M^z\,{=}\,S$. 

The same procedure employed in Section~\ref{ss.eaAFM} leads to $\H^{(0)}=-(N{-}1)\mu-h$ and Eqs.~\eqref{e.H2} and~\eqref{e.H4} turn into
\begin{eqnarray}
 \H^{(2)} &=& \sum_{i=1}^{N-1}\big[
      p_ip_{i+1}{-}q_iq_{i+1}+2\mu(r_i{+}r_{i+1}) \big]
 \notag\\ 
      && \hspace{30mm} - 2h \sum_{i=1}^N (-)^i r_i ~,
\label{e.H2chain}
\\
 \H^{(4)} &=& \frac12\sum_{i=1}^{N-1}\big[
      (r_i{+}r_{i+1}) (q_iq_{i+1}{-}p_ip_{i+1})
 \notag\\ && \hspace{30mm} -2\mu r_ir_{i+1} \big] ~.
\label{e.H4chain}
\end{eqnarray}

\subsection{Quadratic part: the eigenvalue problem}
\label{ss.chainquadratic}

Making $r_j$ explicit, the quadratic Hamiltonian~\eqref{e.H2chain} becomes identical to~\eqref{e.H2a}, with $\bX=0$ and
\begin{eqnarray}
 \bA^2 &\equiv& \mu \bM+h\bH+\bK
\notag\\
 \bB^2 &\equiv& \mu\bM+h\bH-\bK ~,
\end{eqnarray}
where the matrices $\bM$, $\bH$, and $\bK$ are given by
\begin{eqnarray}
 M_{ij}&=&\delta_{ij}(2-\delta_{i1}-\delta_{iN})
\notag\\
 H_{ij}&=&(-)^{i-1}\delta_{ij}
\notag\\
 K_{ij}&=&\delta_{i,j+1}+\delta_{i,j-1} ~;
\label{ee.MHK}
\end{eqnarray}
note that $\bH^2=\mathds{1}$ and
\begin{equation}
 \bH\bM\bH=\bM
~,~~~~
 \bH\bK\bH=-\bK ~,
\label{e.symmetry}
\end{equation}
so it follows that 
\begin{equation}
 \bH\bA^2\bH=\bB^2
~,~~~~~~
 \bH\bB^2\bH=\bA^2 ~.
\label{ee.HA=BH}
\end{equation}
From Eq.~\eqref{e.diagA2} one can see that $\bB^2$ is diagonalized by the orthogonal matrix $\bV\bH$ and has the same eigenvalues of $\bA^2$.
The diagonalization~\eqref{e.diagAB2A} that gives the true frequencies of the system can be simplified observing that
\begin{equation}
 \bA\bB^2\!\bA = (\bA\bH\bA)^2 ~,
\label{e.AHA2}
\end{equation}
and hence it is sufficient to diagonalize the matrix $\bA\bH\bA$, whose components are
$(\bA\bH\bA)_{ij}\,{=}\,\sum_\ell(-)^{\ell{-}1}A_{i\ell}A_{\ell{j}}$,
\begin{equation}
 \bO (\bA\bH\bA) \bO^t =  \tilde\bOm \equiv{\rm diag}\{\tilde\omega_k\} ~.
\label{e.diagAHA}
\end{equation}
Note that
\begin{equation}
 {\rm{Tr}}(\bA\bH\bA)={\rm{Tr}}(\bH\bA^2)  =Nh ~,
\label{e.trAHA}
\end{equation}
since the diagonal of $\bH\bM$ is $(1,-2,2,...,-2,1)$; the positive square-root of $\bA\bB^2\!\bA$ is $|\bA\bH\bA|=\bO^t\bOm\,\bO$, i.e., the matrix whose eigenvalues are $\{\omega_k\,{=}\,|\tilde\omega_k|\}$.
Exchanging $\bA^2$ with $\bB^2$ is equivalent to exchanging $\bq$ with $\bp$, so the correlation matrices share the property \eqref{ee.HA=BH}, i.e., 
\begin{equation}
 \ave{\bp\bp^t} = \bH\ave{\bq\bq^t}\bH~;
\label{e.HqqH=pp}
\end{equation}
in turn, these imply the symmetry relations~\eqref{e.corrsymm}.

Remarkably, the eigenvalues $\omega_k$ are linear in the field $h$, while the zero-$T$ correlators even do not depend on it. Indeed, assume that there is a field interval $h_-<h<h_+$ where $\bA^2$ is positive definite, in such a way that one can calculate $\bA$ and the eigenvalues of $\bA\bH\bA$. Multiplying Eq.~\eqref{e.diagAHA} by $\bG\,{\equiv}\,\bA\bO^t$ the eigenvalue equation becomes
\begin{equation}
 \bG\tilde\bOm = (\bA^2\bH)\,\bG=(h\mathds{1}+\mu\bM\bH+\bK\bH)\,\bG ~.
\label{e.eigenG}
\end{equation}
Therefore the eigenvalues are {\em linear} in $h$, $\tilde\omega_k(\mu,h)=\tilde\omega_{0k}(\mu)\,{+}\,h$, in agreement with the trace condition~\eqref{e.trAHA}, which also entails that $\sum_k\tilde\omega_{0k}\,{=}\,0$. It turns out that the largest negative eigenvalue $\tilde\omega_{k+}$ and the smallest positive one $\tilde\omega_{k-}$ are related to the upper and lower critical fields $\tilde\omega_{k_+}\,{\equiv}\,h{-}h_+<0$ and $\tilde\omega_{k_-}\,{\equiv}\,h{-}h_->0$, since they correspond to the smallest eigenvalues of $|\bA\bH\bA|$, $\omega_{k_+}\,{\equiv}\,h_+{-}h$ and $\omega_{k_-}\,{\equiv}\,h{-}h_-$. When $h$ reaches the borders of the interval $(h_-,h_+)$ the AP configuration becomes unstable, and when $h$ varies within the stability interval there are no sign changes of the eigenvalues $\tilde\omega_k$, so $\varepsilon_k\equiv\tilde\omega_k/\omega_k$ keeps being either $+1$ or $-1$, i.e., the matrix $\bE\,{\equiv}\,{\rm{diag}}\{\varepsilon_k\}$ is fixed ($\bE^2\,{=}\,\mathds{1}$).
Also the (right) eigenvectors $\bm{v}_k$ of Eq.~\eqref{e.eigenG} only depend on $\mu$ and not on $h$, which implies that $\bG$ can depend on $h$ only through  prefactors of its columns, $G_{ik}=f_k(h)~v_{k,i}$. 
Defining the matrix $\bR\equiv\bG\bOm^{-1/2}$, from Eq.~\eqref{e.correl4} the zero-$T$ correlator reads
\begin{equation}
 2\tS\, \ave{\bq\bq^t} = \bR\bR^t
\label{e.qqRR}
\end{equation}
It is easy to obtain the equality $\bR\bE\bR^t=\bH$, which implies that $\bR^{-1}=\bE\bR^t\bH$, and hence $\bR^t\bH\bR=\bE$; the last equality explicitly reads $f_k^2\omega_k^{-1}\sum_i(-)^{i-1}v_{k,i}=\varepsilon_k$, so that the $h$-dependent factors can be taken as $f_k=\omega_k^{1/2}$ and the components $R_{ik}=v_{k,i}$ are in fact independent of $h$, and so are in turn the zero-$T$ correlators $\ave{\bq\bq^t}$ and  $\ave{\bp\bp^t}$, given by Eqs.~\eqref{e.qqRR} and~\eqref{e.HqqH=pp}.

The total magnetization of the AFM chain is
\begin{equation}
 M^z = \tS\sum_i(-)^{i-1}\big[1-\ave{q_iq_i}\big]
     = \tS\big[1-{\rm{Tr}}\big(\bH\ave{\bq\bq^t}\big)\big]
\end{equation}
and
\begin{equation}
 2\tS\, {\rm{Tr}}\big(\bH\ave{\bq\bq^t}\big)
    = {\rm{Tr}}\big(\bH\bR\bR^t\big)
    = {\rm{Tr}}\bm{E}=1 ~,
\end{equation}
since the number of positive and negative
eigenvalues of $\bA\bH\bA$ is, by Sylvester's {\em inertia law}~\cite{Sylvester1852}, the same of those of $\bH$. Eventually, the total magnetization does not change with $h$ and is equal to
\begin{equation}
 M^z = \tS-{\textstyle\frac12} = S ~.
\label{e.Mz}
\end{equation}
Before considering the effect of the quartic interaction, it is to be remarked  that the phase diagram at the HA level can be analytically calculated. Indeed, the AP configuration is stable as long as $\bB^2$ (or, equivalently, $\bA^2$) is positive definite, because if the eigenvalues of $\bB^2$ are all positive then the system's square frequencies $\omega_k^2$, which are the eigenvalues of $\bA\bB^2\bA$, are also positive by the inertia law~\cite{Sylvester1852}. At the phase boundaries $h=h_\pm(N,\mu)$ one of the eigenvalues vanishes, and so the determinant $|\bB^2|=|\bA^2|=0$: the calculation is summarized in Appendix~\ref{a.ha}. The criterion of stability of the minimum of the Hamiltonian corresponds to the classical phase diagram: quantum fluctuations modify it in the way studied in the following Section.

\subsection{Interaction and SCHA}

To quantify the effect of the interaction~\eqref{e.H4chain} the procedure is the same of Section~\ref{ss.AFM.SCHA}, but for the fact that one cannot invoke translation symmetry. The second derivatives are tridiagonal matrices (Appendix~\ref{a.renor}); taking their averages and using the symmetry relations~\eqref{e.symmetry}, which hold true since the trial Hamiltonian satisfies the same symmetry of $\H^{(2)}$, as one can check {\em a posteriori}, the result is
\begin{eqnarray}
 \bigave{\H^{(4)}_{q_iq_i}}_0 &=& \bigave{\H^{(4)}_{p_ip_i}}_0
\notag\\
  &=& \ave{q_iq_{i+1}}_0{+}\ave{q_iq_{i-1}}_0
     {-}\mu\ave{q_{i+1}^2}_0 {-}\mu\ave{q_{i-1}^2}_0
\nonumber\\
 \bigave{\H^{(4)}_{q_iq_{i\pm1}}}_0 &=& -\bigave{\H^{(4)}_{p_ip_{i\pm1}}}_0
 \notag\\
 &=& {\textstyle\frac12}\ave{q_i^2}_0
 +{\textstyle\frac12}\ave{q_{i\pm1}^2}_0-\mu\ave{q_iq_{i\pm1}}_0 ~.
\label{e.SCHAchain}
\end{eqnarray}
The trial Hamiltonian~\eqref{e.H0} keeps the structure of the quadratic Hamiltonian~\eqref{e.H2chain}; indeed implementing the SCHA amounts to renormalizing the matrices
\begin{eqnarray}
 A^2_{ij} &\longrightarrow& A^2_{ij} + \bigave{\H^{(4)}_{p_ip_j}}_0
\notag\\
 B^2_{ij} &\longrightarrow& B^2_{ij} + \bigave{\H^{(4)}_{q_iq_j}}_0 ~,
\end{eqnarray}
which keep being tridiagonal, their elements being equal on the diagonal and opposite on the second diagonals, such that they can be described with suitably modified $\bM$ (diagonal part) and $\bK$ (hopping part), which lose  uniformity (though mirror symmetry is preserved), while the field term $h\bH$ is unchanged. The relations~\eqref{e.symmetry} for the renormalized matrices and the conclusions of Section~\ref{ss.chainquadratic} keep holding. This fact greatly simplifies the calculations: indeed, one has to find the self-consistent correlators for one field value only, because changing $h$ only modifies the eigenvalues $\tilde\omega_k\,{=}\,\tilde\omega_{0k}\,{+}\,h$.

The renormalized frequencies and correlators can be numerically obtained with the method of section~\ref{ss.chainquadratic}. The self-consistent scheme is implemented by iterating the procedure and recalculating the corrections~\eqref{e.SCHAchain}, i.e., starting from Eqs.~\eqref{e.qqRR} and~\eqref{e.HqqH=pp} as estimates of the quantum correlators yielding the renormalized $\bA^2$ and $\bB^2$, which in turn are used to recalculate frequencies and correlators again, until the results are stable.

One can give an analytical estimate of the importance of quantum fluctuations by considering the isotropic ($\mu\,{=}\,1$) and translation-invariant limit, for which  Eqs.~\eqref{e.SCHAchain} become
\begin{equation}
 \ave{\H^{(4)}_{q_iq_i}}_0 = - 2{\cal{D}}
~,~~~~~
 \ave{\H^{(4)}_{q_iq_{i\pm1}}}_0 = {\cal{D}} ~,
\end{equation}
where ${\cal{D}}\equiv{D\,{-}\,D'}$ and, at zero-$T$ and $h\,{=}\,0$, using Eq.~\eqref{e.DD1AFM} it is
\begin{equation}
 {\cal{D}} 
 = \frac1{2\tS N} \sum_k \sqrt{1-\cos^2\!{k}} = \frac1{\pi\tS} ~,
\label{ee.Dtranslinv}
\end{equation}
which is also a useful benchmark of the numerical outcomes when considering the finite chain.

The numerical iterative procedure has been applied in order find the quantum renormalized upper and lower critical fields $h_+(\mu,S)$ and $h_-(\mu,S)$ for different spin values~\cite{footnote3} reported in Ref.~\onlinecite{CPV2014}. At variance with the  na\"{\i}ve expectation, after the result of Section~\ref{s.spinflopbulk}, that quantum fluctuation acted to weaken the ordered phase, quite unexpectedly it happens that in this finite-size system they turn out to {\em stabilize} the AP phase, i.e. $h_+(\mu,S)>h_+(\mu,\infty)$. As a matter of fact, a different mechanism drives the transition: the AP phase is not stabilized by the exchange anisotropy, as in the bulk AFM, but by the field itself, i.e., by the Zeeman energy gain from the unbalanced spin. 
Note that, in spite of this `classical' language, there is no easy way to characterize the quantum SF state of the finite chain.

\section{Conclusions}
\label{s.conclusions}

The first part of this paper is concerned with the proposal of using a Weyl symbol approach for systems which are customarily treated in terms of Boson creation/annihilation operators in normal ordering. Advantages of Weyl symbols are in their more intuitive semiclassical interpretation and in the fact that they are real for Hermitian operators. Useful identities have been recalled and new ones have been derived in order to `convert' from the Boson operator formalism. It is also argued that the phase-space formalism, which is naturally preferred in the Weyl symbol approach, allows for easier and more transparent linear canonical transformations, such that the self-consistent harmonic approximation (SCHA) is introduced in a natural way. This is expected to be useful especially for nanosystems, for which one cannot invoke translation-symmetry to simplify the reduction of the Hamiltonian to normal form.

The second part of the paper gives two practical examples of the phase-space approach, applied to a bulk and to a finite antiferromagnet, respectively, looking at the effect of quantum fluctuations upon the phase boundary between the antiparallel and the spin-flop phases. After suitable Weyl symbols for the spin operators are introduced, the harmonic approximation and the SCHA accounting for the interactions are illustrated for both systems. The first one allows one to compare with the bosonic approach used in previous literature. The second application illustrates how a typical nanosystem can be conveniently treated: thanks to the phase-space approach it was indeed possible to compute the counterintuitive results reported in Ref.~\onlinecite{CPV2014}.

\newpage

\appendix

\section{Classical phase diagram of the finite AFM chain}
\label{a.ha}

Setting $a=\mu{+}h$, $b=2\mu{-}h$, $c=2\mu{+}h$, the determinant of the $(2M{+}1){\times}(2M{+}1)$ matrix $\bA^2$ is
\begin{equation}
D_{2M{+}1} \equiv
\begin{vmatrix}
	~a &  1 &    &    &    &    &    &    \\
	~1 &  b &  1 &    &    &    &    &    \\
	&  1 &  c &  1 &    &    &    &    \\
	&    &  1 &  b &  1 &    &    &    \\
	&    &    &\ddots&\ddots&\ddots& & \\
	&    &    &    &  1 &  c &  1 &    \\
	&    &    &    &    &  1 &  b & 1~ \\
	&    &    &    &    &    &  1 & a~ \\
\end{vmatrix} ~;
\label{e.D}
\end{equation}
it can be expanded by the first/last row expansion, yielding
\begin{eqnarray}
 bc\,D_{2M{+}1} &=&
		a^2b\,(X_{M}+X_{M-1})-2abc\,X_{M{-}1}
\notag\\
  && \hspace{15mm}		+c\,(X_{M-1}+X_{M-2})~.
\label{e.bcD}
\end{eqnarray}
where the $2M{\times}2M$ determinants
\begin{equation}
 X_M\equiv
 \begin{vmatrix}
 	~b &  1 &    &    &    &    \\
 	~1 &  c &  1 &    &    &    \\
 	&  1 &  b &  1 &    &    \\
 	&    &  1 &  c &    &    \\
 	&    &    &\ddots&\ddots \\
 \end{vmatrix}
\end{equation}
are easily shown to satisfy the recursion relation
\begin{equation}
X_{M{+}1}+X_{M{-}1}=2x\,X_{M}
\end{equation}
with $2x\equiv{bc}{-}2$ and initial conditions $X_0\,{=}\,1$, and $X_1\,{=}\,bc{-}1=2x{+}1$. These recursion relations are those of Chebyshev's polynomials of the first kind $T_M(x)$ and of the second kind $U_M(x)$, which differ for the initial condition $T_1=x$ and $U_1=2x$. Therefore 
\begin{equation}
 X_M = U_M+\frac{U_M-T_M}x
     = \frac{\Im\big\{e^{i(M{+}\frac12)k}\big\}}{\sin\frac{k}2}~
\end{equation}
where as usual $x\equiv\cos{k}$. Inserting this result in Eq.~\eqref{e.bcD} and replacing the expressions of $a$, $b$, and $c$, one finds
\begin{eqnarray}
 \sin{k}~D_{2M{+}1} &=& \Im\big\{ e^{iMk}[(a^2b{-}2a)-(2a{-}c)e^{-ik}]\big\}
\notag\\
 &=&  \rho_k~ \cos\big[(2M{+}1){\textstyle\frac{k}2}+\varphi_k\big]
\label{e.DN}
\end{eqnarray}
where the phase is given by
\begin{equation}
 \tan\varphi_k= \frac1r~\cot{\textstyle\frac{k}2} ~,
\end{equation}
with
\begin{equation}
 r \equiv 1+\frac{2h}{(2\mu{-}h)(1{-}\mu^2)} ~.
\end{equation}
Eq.~\eqref{e.DN} entails that the determinant $D_{2M{+}1}$ vanishes when $N=2M{+}1=(\frac\pi{2}-\varphi_k)/({\textstyle\frac{k}2})$, i.e., for 
\begin{equation}
 N_{\rm{c}}(h,\mu) = \frac{\tan^{-1}\!rz}{\tan^{-1}\!z}
\end{equation}
with
\begin{equation}
 z^2 = \tan^2\frac{k}2 = \frac4{bc}-1
  =\frac{1{-}\mu^2{+}h^2/4}{\mu^2{-}h^2/4} ~.
\end{equation}

\section{Second derivatives of $\H^{(4)}$}
\label{a.renor}

With the assumption that any variable with label $0$ or $N{+}1$ vanishes, from Eq.~\eqref{e.H4chain} one finds the first derivatives
\begin{eqnarray*}
 8\, \H^{(4)}_{p_i}&=&
 -3p_i^2(p_{i-1}{+}p_{i+1})
 -(p_{i-1}^3{+}p_{i+1}^3)
\notag\\ && 
 +2p_iq_i(q_{i-1}{+}q_{i+1})
 -q_i^2(p_{i-1}{+}p_{i+1})
 -q_{i-1}^2p_{i-1}
\notag\\ && 
 -q_{i+1}^2p_{i+1}
 -4\mu p_i(p_{i-1}^2{+}p_{i+1}^2+q_{i-1}^2{+}q_{i+1}^2 )
\\
 8\, \H^{(4)}_{q_i}&=&
 3q_i^2(q_{i-1}{+}q_{i+1})
 +(q_{i-1}^3{+}q_{i+1}^3)
\notag\\ &&
 -2q_ip_i(p_{i-1}{+}p_{i+1})
 +p_i^2(q_{i-1}{+}q_{i+1})
 +p_{i-1}^2q_{i-1}
\notag\\ &&
 +p_{i+1}^2q_{i+1}
 -4\mu q_i(q_{i-1}^2{+}q_{i+1}^2+p_{i-1}^2{+}p_{i+1}^2)
\end{eqnarray*}
and the second derivatives
\begin{eqnarray*}
4\, \H^{(4)}_{p_ip_i}&=&  -3p_i(p_{i-1}{+}p_{i+1}) +q_i(q_{i-1}{+}q_{i+1})
\notag\\ &&
-2\mu (p_{i-1}^2{+}p_{i+1}^2{+}q_{i-1}^2{+}q_{i+1}^2)
\notag\\
8\, \H^{(4)}_{p_ip_{i+1}}&=&
-3p_i^2-3p_{i+1}^2-q_i^2-q_{i+1}^2  -8\mu p_ip_{i+1}
\nonumber\\
4\, \H^{(4)}_{q_iq_i}&=&  3q_i(q_{i-1}{+}q_{i+1}) -p_i(p_{i-1}{+}p_{i+1})
\notag\\ &&
-2\mu (q_{i-1}^2{+}q_{i+1}^2{+}p_{i-1}^2{+}p_{i+1}^2)
\nonumber\\
8\, \H^{(4)}_{q_iq_{i+1}}&=&
3q_i^2+3q_{i+1}^2+p_i^2+p_{i+1}^2 -8\mu q_iq_{i+1} ~.
\end{eqnarray*}
Taking their averages one finds Eqs.~\eqref{e.SCHAchain}.


\begin{thebibliography}{00}
\bibitem{footnote1}
 If $\bA^2$ is a multiple of $\mathds{1}$, the matrix $\bX$ must be symmetric; for instance, a particle in a magnetic field entails an antisymmetric $\bX$ and such a system cannot be reduced to the harmonic-oscillator form.
\bibitem{footnote2}
 Due to the existence of two `spin-flop' fields, we use here the subscripts $\pm$ rather than $SF$.
\bibitem{footnote3}
 In the numerical implementation a difficulty arises in the easy-plane case when $N$ is beyond the largest classical value where an AP interval $[h_-,h_+]$ exists, because the bare matrices $\bA^2$ and $\bB^2$ are not positive definite. This can be overcome starting with $\mu=1$ and progressively decreasing $\mu$ while maintaining the correlators. The magnetic field in the subsequent iterations it is kept within the AP interval.
\end{thebibliography}
\end{document}